\documentclass[twocolumn]{aastex631}
\usepackage[T1]{fontenc}
\usepackage{ae,aecompl}
\usepackage{graphicx}
\usepackage{amsmath}
\usepackage{amssymb}
\usepackage{supertabular}
\usepackage{xcolor}
\usepackage{academicons}
\usepackage{multirow}
\usepackage{threeparttable} 
\usepackage[normalem]{ulem}  
\usepackage{times}
\usepackage{hyperref}%
\usepackage{csvsimple}

\begin{document}
\title{Timing results of 22 years for PSR J0922+0638}

\author{Peng Liu}
\affiliation{Department of Astronomy, Xiamen University, Xiamen 361005, China; liang@xmu.edu.cn}
\affiliation{Xinjiang Astronomical Observatory, Chinese Academy of Sciences, Urumqi 830011, China; yuanjp@xao.ac.cn}

\author{Mingyang Wang}
\affiliation{Department of Astronomy, Xiamen University, Xiamen 361005, China; liang@xmu.edu.cn}

\author[0000-0002-5381-6498]{Jianping Yuan}
\affiliation{Xinjiang Astronomical Observatory, Chinese Academy of Sciences, Urumqi 830011, China; yuanjp@xao.ac.cn}
\affiliation{State Key Laboratory of Radio Astronomy and Technology, Beijing 100101, China}
\affiliation{Xinjiang Key Laboratory of Radio Astrophysics, 150 Science 1-Street, Urumqi 830011, China}

\author[0000-0001-6836-9339]{Zhonghao Tu}
\affiliation{Department of Astronomy, Xiamen University, Xiamen 361005, China; liang@xmu.edu.cn}

\author[0000-0002-0786-7307]{Ang Li}
\affiliation{Department of Astronomy, Xiamen University, Xiamen 361005, China; liang@xmu.edu.cn}

\author[0000-0003-4686-5977]{Xia Zhou}
\affiliation{Xinjiang Astronomical Observatory, Chinese Academy of Sciences, Urumqi 830011, China; yuanjp@xao.ac.cn}
\affiliation{State Key Laboratory of Radio Astronomy and Technology, Beijing 100101, China}
\affiliation{Xinjiang Key Laboratory of Radio Astrophysics, 150 Science 1-Street, Urumqi 830011, China}

\author{Na Wang}
\affiliation{Xinjiang Astronomical Observatory, Chinese Academy of Sciences, Urumqi 830011, China; yuanjp@xao.ac.cn}
\affiliation{State Key Laboratory of Radio Astronomy and Technology, Beijing 100101, China}
\affiliation{Xinjiang Key Laboratory of Radio Astrophysics, 150 Science 1-Street, Urumqi 830011, China}

\begin{abstract}
We conducted a timing analysis of PSR J0922+0638 (B0919+06) using data from the Nanshan 26 m radio telescope and the MeerKAT telescope, spanning from January 2001 to March 2023. During this 22-year period, we discovered a previously unreported small glitch (glitch 1) before the well-known large glitch (glitch 2), occurring at ${\rm MJD} \sim 53325(3)$, with a frequency jump amplitude of $\Delta \nu/\nu \sim 0.79(6) \times 10^{-9}$.
We also identified ten slow glitch events, half of which were newly detected. These slow glitches occurred quasi-periodically, with an average interval of approximately 553(21) days, fractional frequency changes ranging from $\Delta \nu/\nu \sim 1.13(1) \times 10^{-9}$ to $4.08(5) \times 10^{-9}$, and a maximum fractional change in the first derivative of the frequency of $\Delta \dot{\nu}/\dot{\nu} \sim -4.6 \times 10^{-3}$.
Additionally, our timing noise analysis reveals a change in the spectral index for noise power before and after glitch 2, with values of $-6.0$ and $-5.3$, respectively, likely due to this large glitch.
Throughout the entire observation period, the first derivative of the spin frequency ($\dot{\nu}$) showed a periodic structure. The possible modulation period was estimated to be 537(24) days before the 700-day data gap at MJD 56716 and 600(58) days afterward.
We discuss the periodic oscillations in pulsar rotation as a possible manifestation of spin-down noise and quasi-periodic slow glitches.
\end{abstract}


\section{Introduction} \label{sec:intro}

Pulsars are highly compact, strongly magnetized rotating neutron stars that exhibit high rotational stability while gradually spinning down. 
The spin-down behavior of young pulsars is often disrupted by two primary types of timing irregularities: glitches and timing noise~\citep{HobbsLK2010, HaskellB2015}.
Glitches are sudden, discrete jumps in the pulsar's rotational frequency, while timing noise refers to stochastic fluctuations in the pulse arrival times. Timing noise can appear as ``white'' noise, where the power is evenly distributed across all fluctuation frequencies, or as ``red'' noise, where the timing residuals are dominated by slow, long-term variations.

Red noise over long timescales is believed to primarily originate from irregularities in pulsar rotation. These irregularities can arise from a variety of factors. For instance, PSR B1828$-$11 exhibits periodic variations in its spin-down rate ($\dot{\nu}$, where $\nu$ is the spin frequency) with cycles of approximately 250 and 500 days. These variations are accompanied by changes in the pulsar's pulse profile, which shifts between ``wide'' and ``narrow'' extremes. 
\citet{stairs2000} suggested that free precession of the spin axis could be the cause of the periodic modulation. \citet{LyneHSS2010} proposed that the spin-down and beam-width variations of PSR B1828$-$11 could similarly be explained by a model in which the magnetosphere switches between states in a quasi-periodic pattern. {\citet{stairs2019} revealed that the mode transition rate could plausibly function as an additional parameter governing the chaotic behaviour in B1828$-$11, as proposed earlier by \citet{seymour2013}.}
Apart from PSR B1828$-$11, other pulsars also show similar irregularities in their rotation and corresponding changes in pulse profiles. \citet{LyneHSS2010} has reported the modulation of the spin-down rate derivatives of 17 pulsars, of which 6 exhibited correlations with changes in pulse shape. \citet{Kerr2016} searched for modulation in the timing behaviors of 151 young pulsars, identifying seven cases of periodic variation with timescales of 0.5--1.5 years. 
PSR J0922+0638 (B0919+06), the subject of this study, also exhibits similar spin behavior and has attracted attention due to its distinctive rotational and radiation characteristics.

\begin{table*} 
\small
\centering
\caption{The parameters of PSR J0922+0638, where the number in parentheses represents the uncertainty of the last significant digit for each parameter. 
Columns 1--11 represent the pulsar name (J2000), right ascension (RA), declination (DEC), period ($P$), period derivative ($\dot{P}$), proper motion in RA (PMRA), proper motion in DEC (PMDEC), dispersion measure (DM), surface dipole magnetic field ($B_{\rm s} = 3.2 \times 10^{19} \sqrt{P\dot{P}} $), epoch and characteristic age ($\tau_{\rm c} = P/(2\dot{P})$), respectively.
\label{tab:pa}}    
    \setlength{\tabcolsep}{4pt}  
\begin{tabular}{lcccccccccc}
  \hline   \hline 
PSR &RA &DEC &$P$ &$\dot{P}$ &PMRA &PMDEC &DM &$B_{\rm s}$ &Epoch &Age  \\ 
 &(hh:mm:ss) &($\ast^{\degr}:\ast^{\arcmin}:\ast^{\arcsec}$) &(s) &(10$^{-14}$) 
&(mas/yr) &(mas/yr) &(cm$^{-3}$pc) &($10^{12}$ G) & MJD &(kyr) \\ 
\hline
J0922+0638$^{a}$  &09:22:14.022(4)$^{b}$& +06:38:23.30(17)$^{b}$   &0.43063$^{c}$  &1.37294$^{c}$  &18.8(9)$^{d}$ & 86.4(7)$^{d}$ &27.2986(5)$^{e}$ &2.46  &55140 &497  \\  
\hline     \hline 
\end{tabular}
\begin{tablenotes}
\item[ ] \textit{Note}. References for parameters of these pulsars: 
$^a$ \citep{ManchesterLTD1978}; 
$^b$ \citep{HobbsLKM2004}; 
$^c$ \citep{ShabanovaPL2013}; 
$^d$ \citep{BriskenFGH2003}; 
$^e$ \citep{StovallRBD2015}.
\end{tablenotes} 
\end{table*}

PSR J0922+0638 is an isolated pulsar of medium age with a characteristic age of 495 kyr \citep{ShabanovaPL2013}. It was discovered in the second Molonglo Radio Observatory pulsar survey \citep{ManchesterLTD1978}, and subsequent observations with XMM-Newton have successfully captured its faint X-ray counterpart \citep{PrinzB2015,RigoselliM2018}.
The pulsar has a spin period of 0.43063 s, with a spin-down rate of $1.37294 \times 10^{-14}$ $\rm{s\,s^{-1}}$ \citep{ShabanovaPL2013}, as shown in Table \ref{tab:pa}.
The radiation of PSR J0922+0638 exhibits a rare ``swooshes'' phenomenon, that is, occasionally there are several to dozens of continuous pulses shifting to earlier longitude phases and then returning to their normal longitude phases~\citep{RankinRW2006, PereraSWL2015, HanHPT2016, WahlORW2016, ShaifullahTOV2018, YuPLZ2019, RajwadePSR2021, WangWYW2024}.

Over its observational history, PSR J0922+0638 has been suffered with one glitch event, with fractional glitch sizes is $\Delta \nu/\nu \sim  1256.2(4) \times 10^{-9}$ \citep{Shabanova2010,YuanWLW2013,BasuSAK2022}.
\cite{Shabanova2010} reported the detection of twelve slow glitch events (MJD 54892--55254) in PSR J0922+0638. 
The maximum change in spin frequency in these events ranged from 2.9 to 4.6 nHz, with a frequency rise time of approximately 200 d. 
In addition to the periodic oscillations of $\dot{\nu}$~\citep{LyneHSS2010},
\cite{Shabanova2010} further revealed the unique characteristics of the pulsar's spin frequency $\nu$ $-$ a sawtooth-like periodic structure with a period of approximately 600 days.
It was suggested that this periodic fluctuation may be induced by a series of consecutive slow glitches with similar characteristics, and that under the influence of $\nu$, $\dot{\nu}$ also exhibits periodic oscillations.
Later, $\dot{\nu}$ was observed to vary and oscillate~\citep{PereraSWL2015,ShawSWB2022}, 
indicating a strong correlation between $\dot{\nu}$ and radiation. We will revisit this topic in Sec.~\ref{Spin states}.
 
In this study, we primarily analyze 21 years (2001--2022) of timing observations of PSR J0922+0638 from MJD 51915 to 59756, collected by the Nanshan radio telescope. Additionally, we incorporate timing data from the MeerKAT telescope spanning MJD 58755 to 60034 (2020--2023) to supplement our analysis.
By analyzing 22 years of timing data, we identified two normal glitches—one large and one small—as well as ten slow glitches. Among them, one small normal glitch and five slow glitches were newly detected events.
Since the small glitch occurred before the large one, we designate them as glitch 1 and glitch 2, respectively.

The paper is organized as follows: In Section \ref{sec:style}, we provide a brief overview of the timing observations conducted with the Nanshan radio telescope and the MeerKAT telescope, along with a description of the analysis process for glitches and timing noise. Section \ref{sec:Results} presents the results of our timing analysis, including normal glitches, slow glitches, timing noise, and changes in the spin-down rate. In Section \ref{sec:Discussion}, we discuss these findings in detail, followed by a summary of the main conclusions in Section \ref{sec:Conclusions}.

\section{Observations and Data analysis} \label{sec:style}

\subsection{Nanshan and MeerKAT data}

The Nanshan radio Telescope is a horizontal fully steerable Cassegrain parabolic antenna with a diameter of 26 m. 
It is equipped with four receivers covering five frequency bands: 1.3 cm, 3.6 cm, 6 cm, 13 cm, and 18 cm.
Initially, the telescope utilized a 128-channel analog filter bank (AFB), with each channel having a bandwidth of 2.5 MHz. In January 2010, it was upgraded to a digital filter bank (DFB) with 1024 channels, featuring a center frequency of 1556 MHz, a total bandwidth of 512 MHz, and an effective bandwidth of 320 MHz. 
Pulsar observations were conducted using an 18 cm receiver, operating at a center frequency of 1540 MHz with a bandwidth of 320 MHz, covering a frequency range of 1381.25 MHz to 1701.25 MHz~\citep{WangMZW2001}.
The pulsar observation cadence was at least three times per month, with individual observations lasting four minutes before October 2017 and eight minutes thereafter. It is worth noting that due to equipment upgrades between 2014 and 2016, no observation data was collected during this period.
Pulsar timing observations began in January 2000. Since July 2002, a long-term monitoring program has been in place, regularly observing 300 pulsars \citep{WangMZW2001}, accumulating over 22 years of data. Before July 2002, pulsar observations were carried out using a dual-channel room-temperature receiver. This was later replaced with a cryogenic receiver with a system temperature of 20 K, improving sensitivity to 0.5 mJy~\citep{ZouWWM2004}.
After filtering the observation files for high-quality pulse profiles, we obtained a total of 1142 pulse times-of-arrival (ToAs).

To compensate for the sparse observation of Nanshan data after 2020, we introduced the timing observation data of MeerKAT (MJD 58755--60034).
The MeerKAT radio telescope is located in the Karoo Desert of South Africa and is operated by the South African Radio Astronomy Observatory (SARAO).
Its core component is an array of 64 Gregorian antennas with a diameter of 13.5 m, and the maximum baseline reaches 8 kilometers \citep{CarliLSB2024}.
MeerKAT is equipped with the innovative Precise Time Manager (PTM) with a timing accuracy better than 5 ns, providing extraordinary opportunities for the observation of millisecond pulsars.
Currently, the telescope uses L-band receivers with an operating frequencies ranging from 856 MHz to 1712 MHz, with 1024 frequency channels, and may deploy S-band (1.75--3.5 GHz) receivers in the future \citep{BailesJAB2020}.
Meanwhile, polyphase filter bank (PFBs) are used to record the data.
Based on the Thousand-Pulsar-Array (TPA) programme, we know that MeerKAT has been conducting long-term and continuous monitoring of approximately 500 pulsars \citep{JohnstonKKS2020,Song2021WKJ}.
The observation rhythm is generally maintained at least once a month, and the integration time for each observation usually does not exceed 2 min, with sub-integration time is 8 s \citep{KeithJKW2024}.
In particular, from March 2021 to February 2022, the observation interval for 294 pulsars was 1 year, and from November 2019 to May 2021, the observation interval for 74 pulsars was about 200 d \citep{KeithJKW2024}.
MeerKAT timing observation data of PSR J0922+0638 from September 2019 to March 2023, which includes 8 different center frequencies (from 944 MHz to 1628 MHz), resulting in a total of 232 ToAs.

\subsection{Timing Analysis}
\label{sec:Timing Analysis}

We used the pulsar preprocessing software package \texttt{PSRCHIVE} to perform offline processing of the observation data, such as dedispersion and folding subintegrated profiles to obtain the average integrated pulse profile \citep{HotanSM2004,StratenDO2012}. 
To obtain the standard profile with the highest signal-to-noise ratio, \texttt{PSRADD} was used to add the average pulse profiles of all.
Next, the PAT tool of \texttt{PSRCHIVE} was used to cross-correlate this standard profile with each profile to obtain the ToAs relative to the observation site, which was achieved through the Fourier-domain Markov-chain method \citep{HotanSM2004}. 
The uncertainty of ToAs also needs to be rescaled, as detailed in the following
paragraph.
It should be noted that these local ToAs need to
be converted to the Solar system barycenter (SSB) to correct
for the effects of Earth motion. 
This conversion was performed based on the Jet Propulsion Laboratory’s (JPL) planetary ephemeris DE440 \citep{ParkFWB2021}  and the Barycentric Coordinate Time (TCB).

Since the uncertainties in ToA measurements are often underestimated, it is necessary to introduce additional noise correction factors: 
\texttt{EQUAD}, which accounts for extra white noise, and \texttt{EFAC}, which compensates for instrumental distortions \citep{LentatiAHP2014}. 
The corresponding correction formula is ${\delta}_{\rm s}^{\rm 2} = {\rm EFAC}^{2} \times(\delta^{2}+{\rm EQUAD}^{2})$, where $\delta$ is the initial uncertainty of ToAs, and ${\delta}_{\rm s}$ is its new value.
These parameters were determined using the \texttt{EFACEQUAD} plug-in of \texttt{TEMPO2} \citep{EdwardsHM2006,HobbsEM2006}, ensuring that the reduced chi-square values for each fitting region fall within the range of 0.95--1.05 after correction.
For further details, see our previous works, e.g., \citet{LiuYGYZ2024,LiuYGYZ2025}.

We used \texttt{TEMPO2} to fit the ToA data and obtained an accurate rotation phase model. The phase was expanded using a Taylor series \citep{EdwardsHM2006}: 
\begin{equation}
\label{equ:1} 
\phi(t) = \phi_{\rm 0} + \nu(t - t_{\rm 0}) + \frac{\dot{\nu}}{2} (t - t_{\rm 0})^{2} + \frac{\ddot{\nu}}{6}(t - t_{\rm 0})^{3} \ ,
\end{equation}
where $\phi_{\rm 0}$ is the pulse phase at reference epoch $t_{\rm 0}$. 
$\nu$, $\dot{\nu}$, $\ddot{\nu}$ are the pulsar spin frequency and its derivatives, respectively.
The difference between the observed ToAs and those predicted by the spin model is referred to as the timing residuals. 
In most cases, Equation (\ref{equ:1}) accurately describes the pulsar’s rotational behavior, causing the timing residuals to fluctuate around zero.
However, long-term timing observations have revealed irregularities in the spin evolution of many pulsars, which are primarily attributed to glitches and timing noise. These factors will be discussed in the following sections.

\begin{figure*}[ht!]
\centering
\includegraphics[width=18 cm]{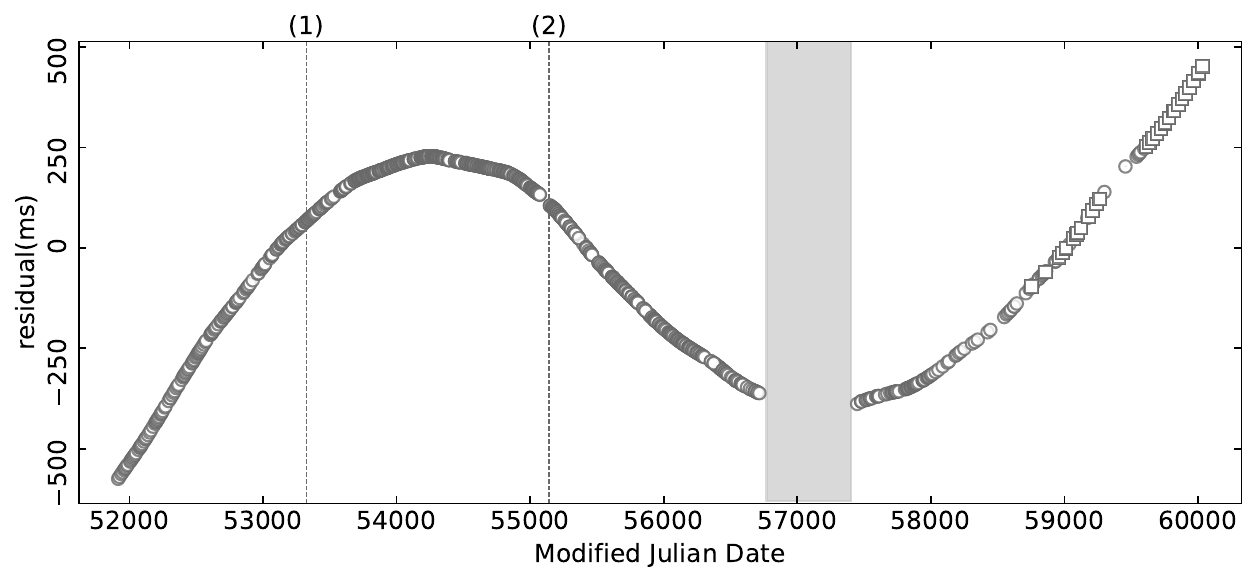}
\caption{
Timing residuals of PSR J0922+0638 relative to the spin-down model and the two glitch models, with the glitch model parameters detailed in Table \ref{tab:gl}.
The hollow circles and solid circles represent the observation data of the AFB and DFB terminals of the Nanshan telescope, respectively, while hollow squares denote the observation data of MeerKAT telescope. 
The vertical dashed line and the numbers in the top brackets indicate the normal glitch epoch and glitch sequence numbers, respectively.
The gray-shaded area represents the gap in timing data due to upgrades at the Nanshan telescope between February 2014 and March 2016 (MJD 56716--57449). 
\label{residual} 
}
\end{figure*}

\subsection{Glitch model}
\label{sec:glitch}

A glitch refers to a sudden jump in the spin frequency of a pulsar on an extremely short timescale~\citep{RadhakrishnanM1969,ReichleyD1969}, offering valuable insights into the structure and internal dynamics of neutron stars~\citep{AntonopoulouHE2022}.
Glitches are still relatively rare, with nearly 700 events detected across over 200 pulsars~\citep{ManchesterHBC2013,BasuSAK2022}.
These glitches come in various forms, including common normal glitches, as well as rarer types such as delayed spin-ups, anti-glitches, and slow glitches~\citep{ZouWWM2004,ArchibaldKNG2013,ShawKLM2021,ZhouGYM2022}.
Moreover, glitches can influence the pulsar's emission properties, such as its pulse profile and spectral characteristics~(e.g. \citealt{WeltevredeJE2011, KeithSJ2013,KouYWY2018,LiuWYS2021,LiuWSY2022,ZhouGYG2023,LiuYGYZ2024,LiuYGYZ2025}).

Typical glitches generally show an exponential recovery followed by a linear decay in the spin-down rate. If the post-glitch recovery phase only involves linear relaxation, the rotation phase model can be corrected by accounting for the permanent changes in the spin phase, frequency, and first derivative of the frequency induced by the glitch~\citep{EdwardsHM2006,YuMHJ2013}:
\begin{equation}
\label{equ:2} 
\phi_{\rm g} = \Delta\phi+ \Delta\nu(t - t_{\rm g}) +   \frac{1} {2} \Delta\dot{\nu} (t - t_{\rm g})^{2}  \ .
\end{equation}
Here, $t_{\rm g}$ represents the epoch of the glitch, and $\Delta\phi$ denotes the permanent change in the pulse phase. The terms $\Delta\nu$ and $\Delta\dot{\nu}$ represent the changes in spin frequency and the first derivative of the frequency, respectively. These changes can also be expressed as fractional glitch sizes, specifically ${\Delta\nu}/{\nu}$ and ${\Delta\dot{\nu}}/{\dot{\nu}}$.


\subsection{Timing noise model}
\label{sec:Timing noise}

Timing noise primarily consists of long-term, continuous noise, often referred to as red noise, which manifests as low-frequency fluctuations in the timing residuals. \cite{BoyntonGHN1972} suggested that timing noise could be described as random walks of pulse phase, spin frequency, and spin-down rate, resulting in corresponding phase noise, spin noise, and spin-down noise.
However, further research found that the random walk model alone is insufficient to explain the complexity observed in the timing residuals \citep{CordesD1985, LyneHSS2010}. According to \cite{ColesHCM2011}, the spectrum of timing noise can be represented by a power-law model, given by:
\begin{equation}
\label{equ:5} 
P(f) = \frac{A}{[1+(f/f_{\rm c})^{2}]^{(\alpha/2)}}\ .
\end{equation}
In this equation, $f$ denotes the frequency of the power spectrum, while $f_{\rm c}$ is the corner frequency, with $f_{\rm c} > 1/T$ ($T$ is the time span of the observed data). Additionally, $A$ is the amplitude, and $\alpha$ is the spectral index. The power law model has been widely adopted in timing noise analysis, yielding significant results in the field \citep{HobbsLK2010, ShannonLKJ2016, ParthasarathySJL2019, EspinozaADS2021, YuanWDK2024}.

\begin{table*} [ht]
\normalsize
\begin{minipage}[]{90mm}
\caption{\raggedright Pre- and post-glitch timing solutions for PSR J0922+0638. 
\label{tab:int}} 
\end{minipage}
\vspace{0.2cm}
\begin{flushleft}  
\hspace*{-\columnsep}  
  \begin{threeparttable}
    \renewcommand{\arraystretch}{1.1}
    \setlength{\tabcolsep}{10pt}    
    \resizebox{1.05 \textwidth}{!}{
\begin{tabular}{lccc}
  \hline   \hline
Parameter          & Pre-glitch 1  & Post-glitch 1  & Post-glitch 2 \\
  \hline
PSR      &  \multicolumn{3}{c}{ J0922+0638} \\
  \hline
Right ascension (J2000) (h:m:s)  &09:22:14.037(4)  &09:22:14.043(4)  &09:22:14.054(4)\\
Declination (J2000) (d:m:s)     &+06:38:24.34(17)  &+06:38:24.72(17) &+06:38:25.52(17)\\
Pulse frequency, $\nu$ (Hz)     &2.32220791446(2)  &2.32219761985(1)  & 2.322179074696(9) \\
First derivative of pulse frequency, $\dot{\nu}$ ($10^{-14}$\rm\ s$^{-2}$) &$-$7.3890(4)   &$-$7.38863(5)   &$-$7.39821(2)  \\
Epoch of frequency determination (MJD)  &52618    &54231    &57591  \\
ToA numbers                             &161      &413      &800      \\
RMS timing residual ($\mu$s)     &4067     &3554     &13379    \\
Data span (MJD)  & 51915--53323       & 53328--55136    & 55149--60034  \\
  \hline
Proper-motion RA     &  \multicolumn{3}{c}{18.8(9)}   \\
Proper-motion DEC    &  \multicolumn{3}{c}{86.4(7)}   \\
Time units           &  \multicolumn{3}{c}{TT(TCB)}   \\
Reference time scale &  \multicolumn{3}{c}{TT(TAI)}   \\
Solar System ephemeris model &  \multicolumn{3}{c}{DE440}   \\
  \hline   \hline
\end{tabular}
}
\end{threeparttable}
\end{flushleft}%
\vspace{-0.2cm}
\end{table*}
\begin{table*}[ht]
\normalsize
\begin{minipage}[]{110mm}
\caption{The timing solutions and glitch parameters obtained for PSR J0922+0638. 
\label{tab:gl}} 
\end{minipage}
\vspace{0.2cm}
\begin{flushleft}
\hspace*{-\columnsep}  
\begin{threeparttable}
    \renewcommand{\arraystretch}{1.1} 
    \setlength{\tabcolsep}{20pt}  
    \resizebox{1.05 \textwidth}{!}{
    \begin{tabular}{lcc}
      \hline \hline
      Parameter             & Glitch 1   &  Glitch 2\\
      \hline
      Pulse frequency, $\nu$ (Hz)            & 2.3222034659(9)  & 2.3221927117(5)  \\
      First derivative of pulse frequency, $\dot{\nu}$ ($10^{-14}$\rm\ s$^{-2}$)  
                            & $-$7.401(2)   & $-$7.388(1) \\
      Epoch of frequency determination (MJD)     & 53315     & 55000      \\
      Data span (MJD)       & 53191--53484  & 54836--55451  \\
      ToA numbers           & 48     & 167      \\
      RMS timing residual ($\mu {\rm s}$)     
                            & 175       & 700         \\
      Glitch epoch (MJD)    & 53325(3)   & 55142(7)  \\
      $\Delta\nu/\nu$ ($10^{-9}$)      
                            & 0.79(6)   & 1257.54(8)     \\
      $\Delta\dot{\nu}/\dot{\nu}$ ($10^{-3}$)         
                            & $-$ & $-$1.0(2) \\
      \hline \hline
    \end{tabular}
    }
\end{threeparttable}
\end{flushleft}%
\vspace{-0.2cm}
\end{table*}

\section{Results} \label{sec:Results}

In this section, we present the analysis of the timing data for PSR J0922+0638 from MJD 51915 to 60034. As detailed in Sec.~\ref{sec:Timing Analysis}, we used \texttt{TEMPO2} to fit the ToAs and construct a timing model that includes the spin frequency ($\nu$) and its first derivative ($\dot{\nu}$). 
Due to the discontinuity in the timing residual phase caused by the large glitch (glitch 2), we incorporated relevant parameters for both glitches into the timing model, thereby achieving phase continuity in the timing residuals over the entire time span (2001--2023), as shown in Fig. \ref{residual}. We observe that the timing residuals exhibit strong red noise, which will be further addressed in Sec.~\ref{sec:noise}.

After careful analysis of the timing residuals, we identified one small glitch, one large glitch, and ten slow glitches. The timing models before and after the normal glitches are shown in Table \ref{tab:int}, with errors representing the standard errors (1 $\sigma$) obtained from \texttt{TEMPO2}.
Table \ref{tab:gl} lists the relevant parameters for the two normal glitches, while Table \ref{tab:slow} provides the parameters for the ten slow glitches.
In both Table \ref{tab:gl} and Table \ref{tab:slow}, the glitch epoch is estimated as the central date between the last observation before the glitch and the first observation after it, with the uncertainty determined as half the time interval between these two observations. The other parameters in Table \ref{tab:slow} are derived using the extrapolation method, and the uncertainties are calculated through error propagation. 

\begin{figure}
\centering
\includegraphics[width=1\linewidth]{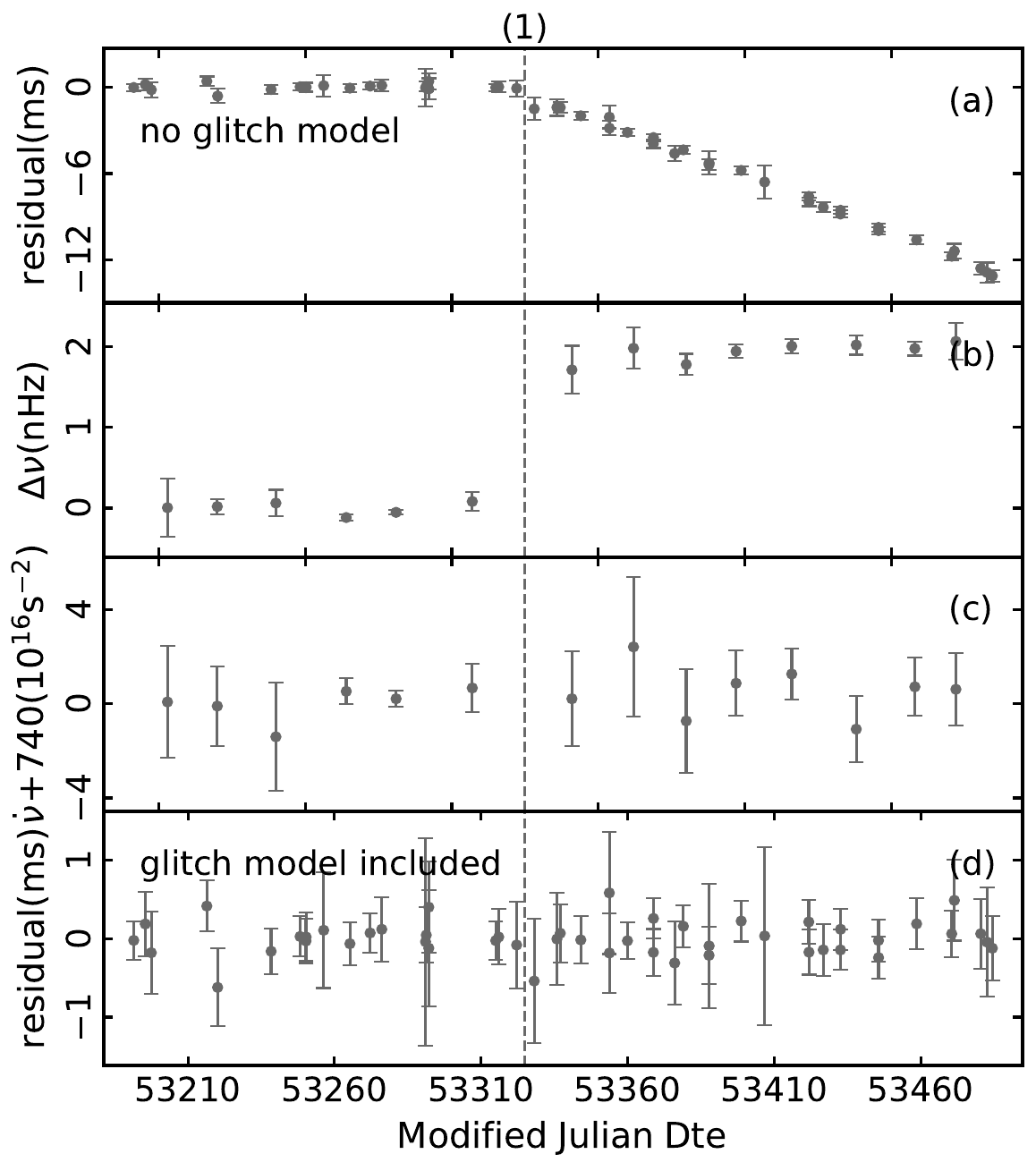}
\caption{Small glitch in PSR J0922+0638: Panel (a) presents the timing residuals obtained by subtracting the pre-glitch rotational model. 
Panel (b) displays the rotational frequency residuals ($\Delta \nu$) relative to the pre-glitch model. The data span used for fitting the glitch parameters corresponds to that shown in panel (a) (MJD 53191--53484). 
Panel (c) depicts the evolution of the spin-down rate ($\dot \nu$) over time after subtracting its average value. Panel (d) shows the timing residuals after incorporating the fitted glitch parameters into the timing model. The vertical dashed line and the numbers in the top brackets indicate the glitch epoch and glitch sequence numbers, respectively.
}
\label{fig:sma}
\end{figure}
%

\subsection{Normal glitches \label{Normal:glitch}}

\cite{Shabanova2010} was the first to report the glitch event (glitch 2) of PSR J0922+0638, which occurred at MJD 55139.8(1) with a glitch size of $\mathbf{\Delta \nu/\nu \sim  1257.1(3)\times 10^{-9}}$ and a corresponding spin-down rate change of $\mathbf{\Delta \dot{\nu}/\dot{\nu} \sim  -7(1) \times 10^{-3}}$.
Subsequently, this glitch was also reported by \cite{YuanWLW2013} and \cite{BasuSAK2022}.

Using timing data from the Nanshan telescope, we also identified this large glitch at MJD 55142(7). Our fitting results indicate a glitch size of $\Delta \nu/\nu \sim  1256.98(9)\times10^{-9}$ with a corresponding spin-down rate change of $\Delta \dot{\nu}/\dot{\nu} \sim  -0.96(22)\times10^{-3}$. 
The fitted timing solutions and glitch parameters are summarized in Table \ref{tab:gl}. Within the range of error, our results are largely consistent with previous studies~\citep{Shabanova2010,YuanWLW2013,BasuSAK2022}. 
Based on observation data of the Nanshan telescope, we did not detect significant changes in the pulse profile of PSR J0922+0638 after the glitches, due to the relatively low signal-to-noise ratio of the pulse profiles.
Previously, it has been reported that, during the ``swooshes'' phenomenon of PSR J0922+0638, the $\sim$ 50 pulse signals exhibited a phase offset of approximately 0.013 cycles~\citep{PereraSWL2015}. 
As the emission shifts occur approximately every 700 rotation periods~\citep{WahlORW2016}, it was estimated that such shifts will result in a change of about 0.00093 cycles in the peak phase of the integrated pulse profile.
Given that the timing white noise residual of most normal pulsars is typically around 0.001 cycles and PSR J0922+0638 is a noisy pulsar with timing maximum residual of $>$1 cycle (Fig.~\ref{residual}), this phase offset has not significantly impacted the timing results.

To search for potential new glitches, we carefully examined the timing residuals before and after the known large glitch. As a result, we identified a characteristic signature of a small glitch (glitch 1) in the timing residuals approximately 200 days before and after MJD 53325(3).
To further investigate this event, we analyzed the evolution of spin frequency and its first derivative, as shown in panels (b) and (c) of Fig. \ref{fig:sma}. In panel (b), a significant jump in $\Delta \nu$ is observed around MJD 53325(3), with an amplitude of approximately 1.8 nHz. However, in panel (c), $\dot{\nu}$ does not exhibit a noticeable jump, which may be attributed to strong red noise.
From Fig. \ref{fig:sma}, it is evident that PSR J0922+0638 experienced a previously unreported small glitch at MJD 53325(3). The estimated glitch size is $\Delta \nu/\nu \sim  0.79(6) \times 10^{-9}$, with detailed parameters provided in Table \ref{tab:gl}. Moreover, the timing residuals relative to the glitch model show only white noise, indicating that the spin behavior has been well characterized, as seen in panel (d) of Fig. \ref{fig:sma}. The consistency between the fitting results and Fig. \ref{fig:sma} further confirms the presence of a small glitch at MJD 53325(3).

To validate the authenticity of this event, we estimated the minimum detectable glitch size for PSR J0922+0638. This analysis follows the detectability limit method proposed by \cite{EspinozaASW2014}, using the equation: 
\begin{equation}
\label{equ:3} 
\Delta{\nu_{\rm lim}} = {\rm max}(\Delta{T}\left| \Delta{\dot{\nu}} \right|/2, (2 \left| \Delta{\dot{\nu}} \right| \sigma_{\rm \phi})^{1/2})\ ,
\end{equation}
where $\Delta{T}$ is the observation cadence, $\Delta{\dot{\nu}}$ is the spin-down rate jump, and $\sigma_{\rm \phi}$ is the typical dispersion of the rotational phase.
For PSR J0922+0638, we use $\Delta{T} = 6$ d, $\left| \Delta{\dot{\nu}} \right| = 3.3 \times 10^{-17} {\rm s}^{-2}$, and $\sigma_{\rm \phi} = 0.01$ rotations.
The results indicate a minimum detectable glitch size of 0.82 nHz, which is significantly smaller than the observed glitch size of 1.84(1) nHz, thereby confirming the validity of this glitch event.

\begin{table*}
\centering
\caption{Parameters of the slow glitches in PSR J0922+0638. 
The first column in the table is the slow glitch number. 
Columns 2--5 indicate the glitch start epoch ($T_{\rm start}$), glitch end epoch ($T_{\rm end}$), spin frequency rise time ($T_{\rm rise}$) and interval to next glitch time ($T_{\rm inter}$), respectively. 
Columns 6 to 7 columns show the maximum change amplitude of frequency ($\Delta\nu_{\rm max}$), and its fractional increase ($\Delta\nu/\nu$). 
Columns 8 to 9 present the maximum change in the first derivative of the spin frequency ($\Delta\dot{\nu}_{\rm max}$) and its fractional change ($\Delta\dot{\nu}/\dot{\nu}$).
Column 10 indicates whether the slow glitch is a newly discovered event, where ``N'' denotes a previously reported event and ``Y'' signifies a new detection.
The final column provides the data span for each glitch.
\label{tab:slow}}    
    \setlength{\tabcolsep}{6.2pt}  
\begin{tabular}{lcccccccccc}
\hline \hline
Gl. No. &  $T_{\rm start}$  &$T_{\rm end}$ &$T_{\rm rise}$ &$T_{\rm inter}$ &  $\Delta \nu_{\rm max}$ &  $\Delta \nu/\nu$  &  $\Delta \dot{\nu}_{\rm max}$ &  $\Delta \dot{\nu}/\dot{\nu}$  &  New?  &  Data span \\
& (MJD)  & (MJD) &(days) &(days) & ($10^{-9}$ Hz)  & (10$^{-9}$)   &(10$^{-17}$ s$^{-2}$)   &(10$^{-3}$) & (Y/N)  &(MJD)\\   
  \hline
8  &52538(8)  &52739(22) &201(23) &494(20) &2.92(7)  &1.26(3)   &36(3)   &$-$4.9(4) &N &52199--53015\\
9  &53032(18) &53246(16) &215(24) &579(20) &5.02(6)  &2.16(2)  &36(4)   &$-$4.9(5) &N &52741--53323\\
10  &53611(9) &53802(17) &191(19) &583(10) &3.17(2)  &1.364(9) &30(3)   &$-$4.1(4) &N &53335--54190\\
11  &54194(4) &54378(25) &184(27) &598(8)  &2.79(4)  &1.20(2)  &29(3)   &$-$4.0(3) &N &53892--54714\\
12  &54792(7) &55005(23) &213(25) &343(7) &4.79(3)  &2.06(2)  &40(2)   &$-$5.4(3) &N &54516--55135\\
13  &55662(5) &55927(16) &265(17) &573(9)  &2.61(3)  &1.13(1)  &12.9(9) &$-$1.7(1) &Y &55325--56228\\ 
14  &56235(6) &56488(14) &253(15) &482(6)  &4.33(3)  &1.86(1)  &32(1)   &$-$4.3(1) &Y &55932--56717\\
15 &58160(24) &58373(21) &213(32) &517(42) &3.74(6)  &1.61(3)  &31(3)   &$-$4.2(4) &Y &57816--58644\\
16 &58677(34) &58923(24) &246(42) &545(36) &4.1(3)  &1.75(12)  &40(7)   &$-$5.4(9)  &Y &58425--59210\\
17 &59222(12) &59569(31) &347(33) &812(12) &9.47(5)  &4.08(2)  &33(2)   &$-$4.5(3)  &Y &58928--60033\\
\hline  \hline
\end{tabular}
\begin{tablenotes}
\item[ ] \textit{Note}. The detailed parameters of slow glitches 1--7 were provided in \cite{Shabanova2010} and are also included in the data file of this work.
\end{tablenotes} 
\end{table*}

\subsection{Slow glitches}
\label{Slow:glitches}

We analyzed the slow glitches in PSR J0922+0638 using timing data spanning from 2001 to 2023. To accurately identify slow glitch events and determine their parameters, we fitted the timing data using a 150–250 day window, obtaining $\nu$ and $\dot{\nu}$ at different epochs.

\begin{figure*}
\centering
\includegraphics[width=18 cm]{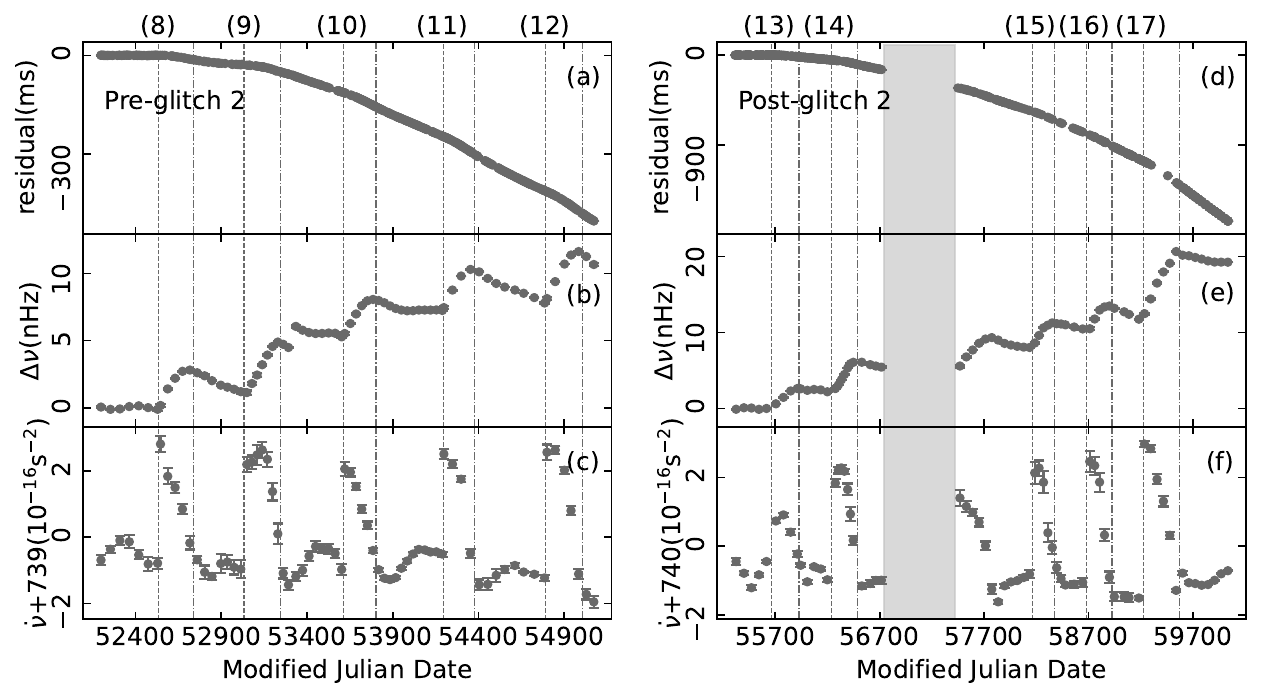}
\caption{Slow glitches in PSR J0922+0638: The left and right panels depict the evolution of $\nu$ and $\dot{\nu}$ before and after glitch 2 (the large glitch). In panel (b) in the left, $\Delta \nu$ exhibits a distinct jump, corresponding to the small normal glitch reported in Sec. \ref{Normal:glitch}.
In the right panels, the gray-shaded region represents a data gap from the Nanshan telescope. Vertical dashed lines mark the start epochs of slow glitches, while vertical dash-dot lines indicate their end epochs. The numbers in the top brackets correspond to the sequence numbers of the slow glitches.
For further details regarding this figure, please refer to panels (a)–(c) of Fig. \ref{fig:sma}. 
\label{glitch:sl}} 
\end{figure*}

In Fig. \ref{glitch:sl}, panels (a) and (b) are plotted relative to the timing behavior before slow glitch 8, while panels (d) and (e) are referenced to the spin trend before slow glitch 13. The figure presents the timing residuals and the evolution of $\nu$ and $\dot{\nu}$, allowing clear identification of ten slow glitches—five previously reported by \cite{Shabanova2010} and five newly discovered in this study. For consistency, we follow the slow glitch numbering system of \cite{Shabanova2010}, labeling the detected events as 8 to 17.

Table \ref{tab:slow} provides updated parameters for slow glitches 8–12 and includes new parameter estimates for slow glitches 13–17. 
The permanent frequency changes ($\Delta\nu_{\rm max}$) of slow glitches 9, 11, and 12 obtained in this study are 5.02(6) nHz, 2.79(4) nHz, and 4.97(3) nHz, respectively, which show significantly discrepancies compared to the results reported by \cite{Shabanova2010}, namely 2.9 nHz, 4.6 nHz, and 3.7 nHz.
This discrepancy likely arises because \cite{Shabanova2010} inferred slow glitch parameters from the overall evolution of $\Delta \nu$ between 1991 and 2009, without individually plotting the slow glitch events.

From Table \ref{tab:slow}, the maximum fractional changes in frequency for these slow glitches range from $\Delta \nu/\nu \sim 1.13(1) \times 10^{-9}$ to $4.08(2) \times 10^{-9}$, with corresponding rise times of $T_{\rm rise} \sim 184 - 347$ days. Among them, slow glitch 17 stands out as the largest slow glitch observed for this pulsar to date, with a maximum frequency change of $\Delta \nu_{\rm max} \sim 9.47(5)$ nHz and the longest rise time of $T_{\rm rise} \sim 347$ days.

Panels (b) and (e) of Fig. \ref{glitch:sl} reveal that $\nu$ increases gradually during a slow glitch and decreases slowly after the glitch ends. Panels (c) and (f) further show that $\dot{\nu}$ exhibits a sudden jump at the beginning of a slow glitch, followed by two distinct evolutionary trends:
\begin{itemize}
    \item Example of slow glitch 8: $\left|\dot{\nu}\right|$ initially decreases to a minimum and then gradually returns to its pre-glitch state;
    \item Example of slow glitch 9: $\left|\dot{\nu}\right|$ drops sharply at the glitch onset, then steadily rises throughout the slow glitch, eventually recovering to its original level.
\end{itemize}

Panels (e) and (f) also indicate that the evolution of $\nu$ and $\dot{\nu}$ between slow glitch 15 and the gray-shaded region follows the characteristics of a slow glitch. However, as it is unclear whether this slow glitch event is complete, its true parameters remain undetermined, and it is not classified as a new slow glitch in this work. Furthermore, based on the periodicity observed in panels (d) and (f), we speculate that an unrecognized slow glitch event may exist within the gray region.

Panels (b), (c), (e), and (f) of Fig. \ref{glitch:sl} highlight the quasi-periodic nature of the slow glitches. From Table \ref{tab:slow}, we find that the average interval between slow glitches is approximately 553(21) days. Interestingly, the maximum fractional change in the first frequency derivative for these slow glitches remains nearly constant at $\Delta \dot{\nu}/\dot{\nu} \sim -4.6 \times 10^{-3}$. The only exception is slow glitch 13, which has a significantly higher value of $-1.7(1) \times 10^{-3}$, possibly due to the influence of glitch 2.

\begin{figure*}[ht!]
\centering
\includegraphics[width=18 cm]{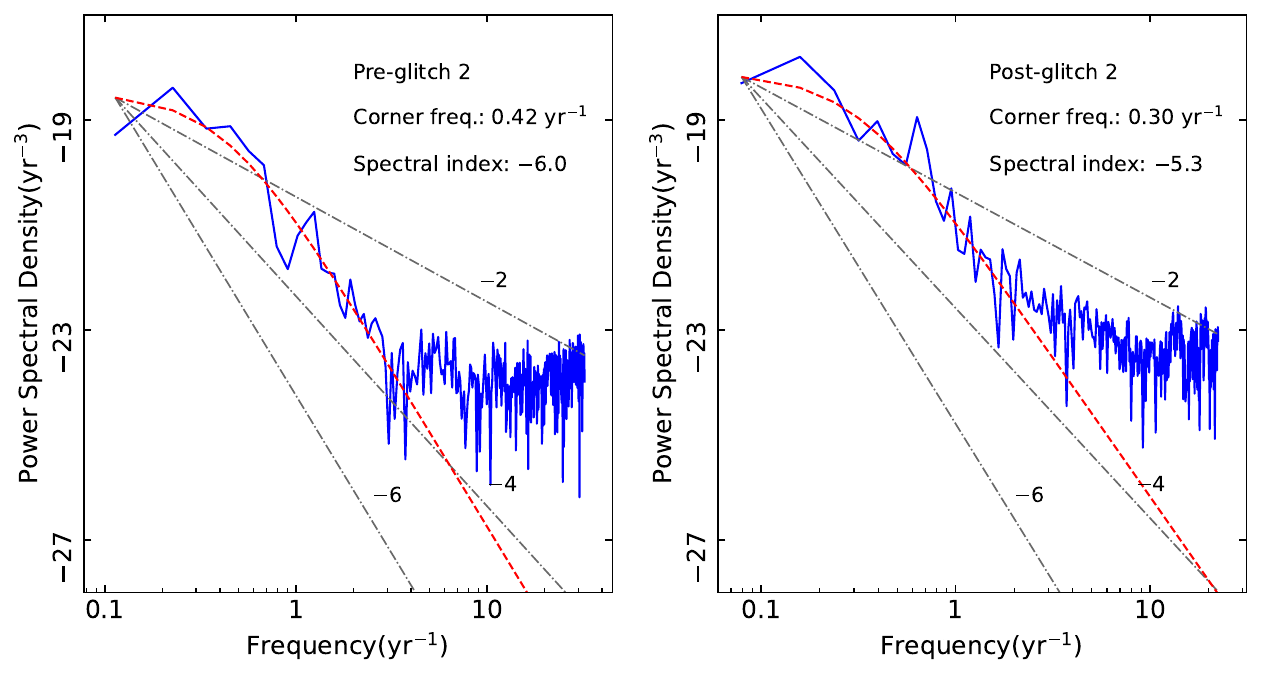}
\caption{Power spectra of timing noise in PSR J0922+0638: The left and right panels illustrate the power spectra of the timing noise before and after glitch 2. The red dashed line represents the best-fit result for the low-frequency portion of the power spectral density (shown as the blue curve). Additionally, the three gray dashed lines correspond to theoretical power-law models with spectral indices of $-2$, $-4$, and $-6$, respectively. These indices, which determine the slope of the PSD, are associated with different noise characteristics: phase noise, spin noise, and spin-down noise \citep{BoyntonGHN1972}.
\label{noise:PSD}}
\end{figure*}   
%
%
\begin{figure*}
\centering
\includegraphics[width=18 cm]{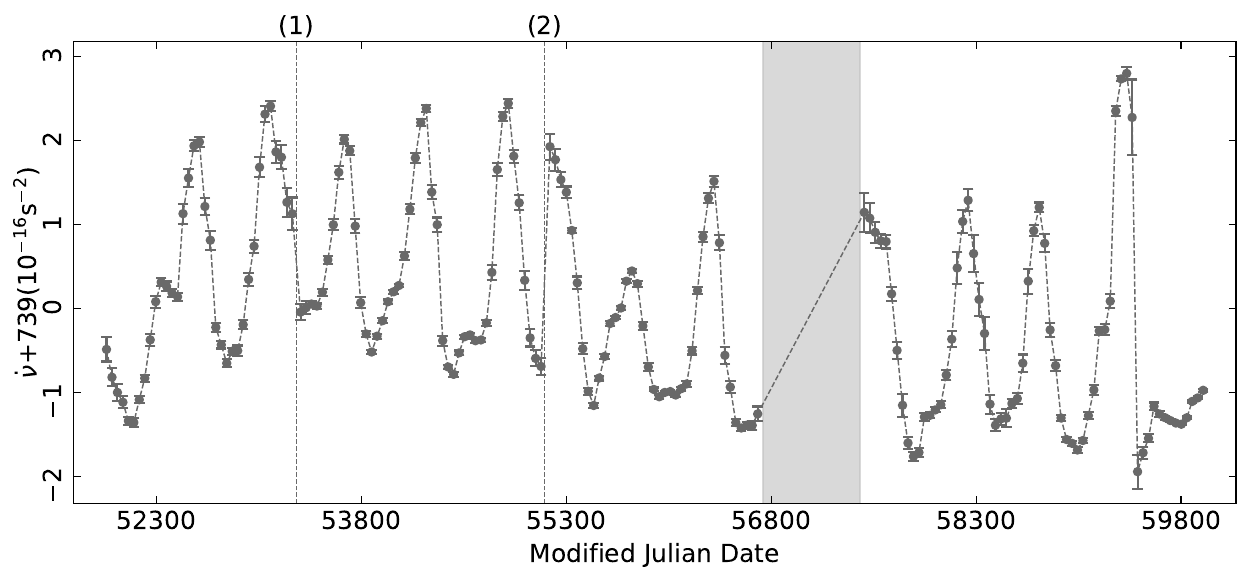}
\caption{Spin evolution behavior of PSR J0922+0638: The vertical dashed line indicates the glitch epoch, and the numbers in the top brackets correspond to the glitch sequence numbers. The gray area represents the data gap from the Nanshan telescope. For further details regarding this panel, please refer to panel (c) of Fig. \ref{fig:sma}.
\label{glitch:n}}
\end{figure*}  
%

\subsection{Timing noise}
\label{sec:noise}

When analyzing the timing data of PSR J0922+0638, we identified significant time-correlated noise (red noise) in the timing residuals. To model this noise and estimate its power spectrum, we employed the ``Cholesky'' method. Specifically, we used the \texttt{SPECTRALMODEL} plug-in of \texttt{TEMPO2} to accurately determine the power spectral density (PSD) of the timing noise. The PSD was then fitted with a power-law model to extract the key parameters. To ensure the accuracy of our analysis, we relied exclusively on Nanshan timing data, as it provides the longest observational span.

Fig.~\ref{noise:PSD} presents the logarithmic relationship between the PSD and frequency of the timing noise. The PSD displays oscillatory characteristics, but shows neither significant peaks nor obvious periodicity. The presence of significant red noise is evident in both the pre- and post-glitch 2 data. Notably, at frequencies approaching 3 $\rm yr^{-1}$, the PSD exhibits a turnover (or reaches white noise levels), indicating that the minimum detectable autocorrelation timescale in the timing data is less than 120 days \citep{AntonelliMAH2023}.

Fitting the power spectrum with a power-law model yielded spectral indices of $-6.0$ and $-5.3$ for the pre- and post-glitch 2 data, respectively. For the pre-glitch 2 segment, the spectral index of $-6.0$ is highly consistent with a random walk in torque model \citep{BoyntonGHN1972}. However, after glitch 2, the spectral index shifts to $-5.3$, suggesting that the timing noise is no longer dominated by a single process but instead results from a combination of angular velocity random walk and torque random walk \citep{AntonelliMAH2023}.

According to \cite{HobbsLK2010}, an analysis of 366 pulsars indicates that timing noise in pulsars with characteristic ages below $10^5$ years is primarily associated with glitch recovery. We speculate that the spectral index change before and after glitch 2 may be closely linked to the glitch itself—possibly triggering alterations in the pulsar's magnetosphere \citep{LyneHSS2010}.

It is also worth noting that across the entire observation span, PSR J0922+0638 consistently exhibits timing behavior influenced by spin-down rate random walk. Given this, we suggest that the periodic oscillations observed in the pulsar's rotational behavior are likely attributable to spin-down noise.

\subsection{Spin-down rate switch}
\label{Spin states}

To investigate the evolution of the spin-down rate ($\dot \nu$) of PSR J0922+0638 over a 22-year period (2001–2023), we applied a spin-down model to fit the ToA data within 300-day windows, with a step size of 40 days. This allowed us to obtain $\dot \nu$ values at different epochs, as shown in Fig. \ref{glitch:n}.

The data clearly reveal that $\dot \nu$ exhibits significant single-peak periodic structures. This finding aligns with previous studies \citep{LyneHSS2010, Shabanova2010, ShabanovaPL2013, PereraSWL2015}, although it is worth noting that \cite{PereraSWL2015} observed a double-peak periodic structure in the $\dot \nu$ data between MJD 48000 and 56500, which contrasts with our results. We attribute this discrepancy to the shorter fitting interval used by \cite{PereraSWL2015} (150 days with a 15-day step size), which was more sensitive to finer structures. In contrast, the Nanshan telescope data, with its larger fitting intervals and step sizes, limits the resolution of such structures.
Additionally, there is a difference between panel (c) of Fig. \ref{glitch:sl} and the 
$\dot{\nu}$ evolution shown in Fig. \ref{glitch:n}, primarily due to the different timescales used in the fitting process for the two datasets.

\begin{figure}
\centering
\includegraphics[width=1\linewidth]{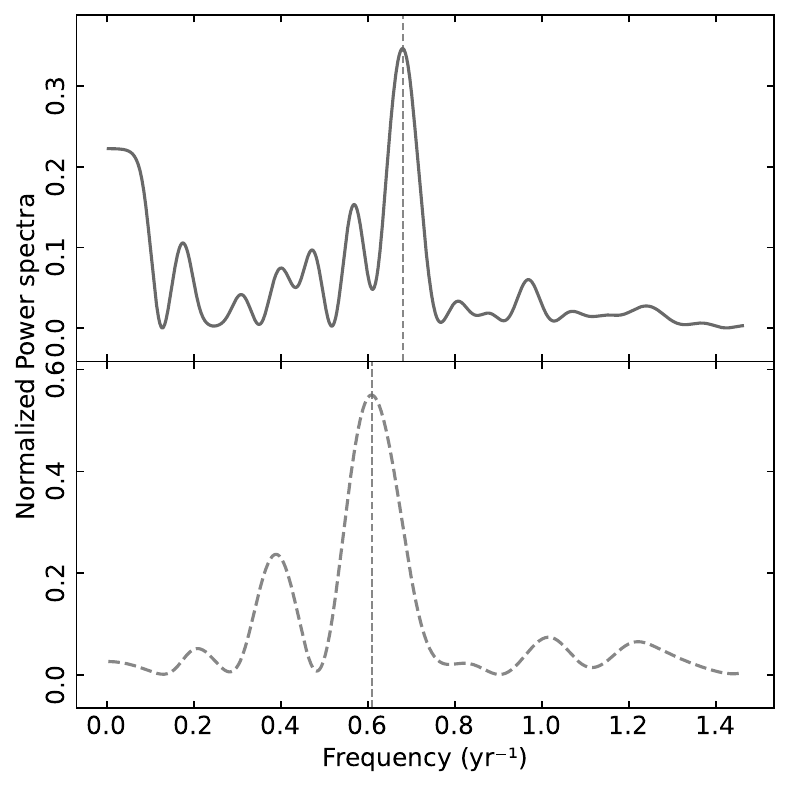}
\caption{Lomb-Scargle periodicity spectra of the $\dot{\nu}$ evolution for PSR J0922+0638. The upper and lower panels display the analysis results for the periods before and after MJD 56716, respectively. The vertical dashed line indicates the frequency corresponding to the peak of the spectrum.
}
\label{0922:lomb}
\end{figure}
\begin{figure}
\centering
\includegraphics[width=1.0\linewidth]{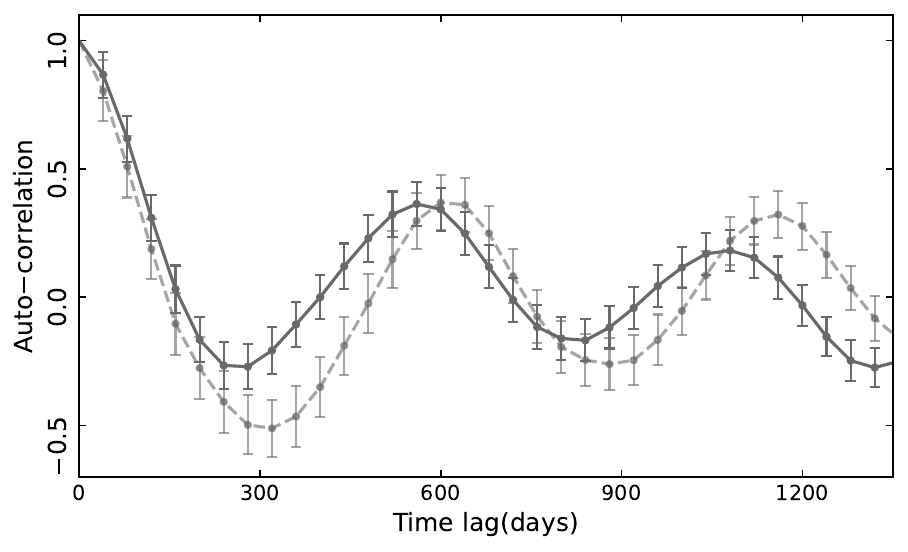}
\caption{The autocorrelation function of $\dot{\nu}$ for PSR J0922+0638 is shown, with the solid line representing the modulation period before MJD 56716 and the dashed line representing the modulation period after MJD 56716.
}
\label{0922:acf}
\end{figure}

From Fig. \ref{glitch:n}, we identify a total of 10 complete periodic structures, each approximately 500--600 days in length. To analyze the periodicity of $\dot{\nu}$, we applied Lomb-Scargle spectral analysis \citep{Scargle1982} to investigate the oscillatory properties of the data. We note that the nearly 700-day data gap may influence the results of this analysis.
To account for this gap, we divided the $\dot{\nu}$ into two independent sets: one before (MJD 51915--56716) and one after (MJD 57448--59755) MJD 56716. The results, shown in Fig.~\ref{0922:lomb}, suggest a possible change in the oscillation frequency of $\dot{\nu}$ before and after MJD 56716.
Specifically, the Lomb-Scargle spectrum before MJD 56716 peaks at 0.68(3) $\rm{yr}^{-1}$, corresponding to a modulation period of approximately 537(24) days.
After MJD 56716, the spectrum peaks at 0.61(6) $\rm{yr}^{-1}$, with a modulation period of approximately 600(58) days. We evaluated the uncertainty of the peak frequency using the method proposed by \citet{ZubietaGEA2024}.

\begin{figure*}[ht!]
\centering\includegraphics[width=17 cm]{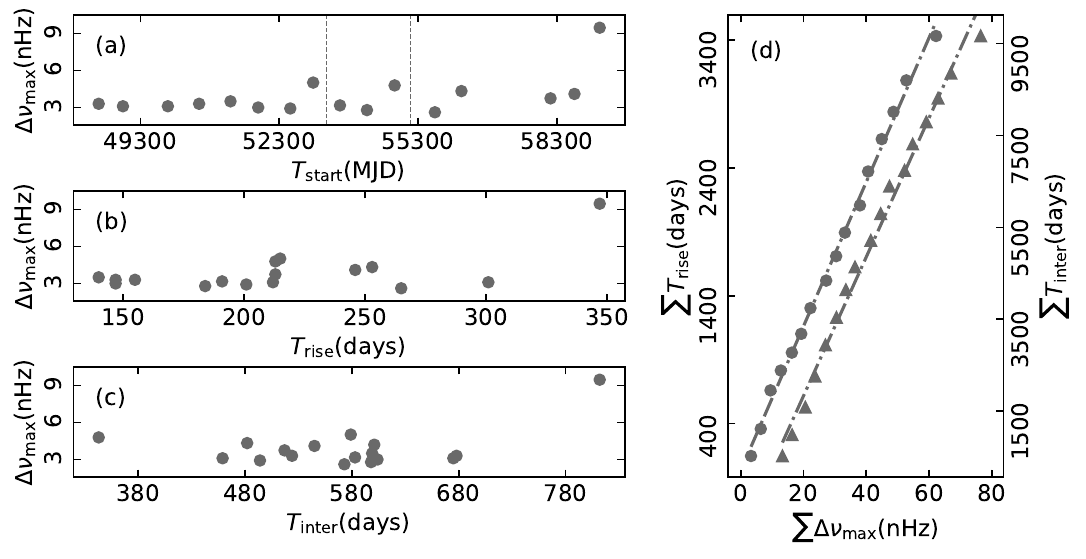}
\caption{Correlations between $\Delta\nu_{\rm max}$ and other parameters for 17 slow glitches in PSR J0922+0638: 
Panels (a), (b), and (c) show the correlation between $\Delta\nu_{\rm max}$ and $T_{\rm start}$, $T_{\rm rise}$, and $T_{\rm inter}$, respectively. In panel (a), the two vertical dashed lines mark the jump epochs of two normal glitches. Panel (d) illustrates the relationship between the cumulative maximum frequency change amplitude $\sum \Delta\nu_{\rm max}$ and the cumulative rise time $\sum T_{\rm rise}$, as well as the cumulative slow glitch interval time $\sum T_{\rm inter}$. Specifically, the circles represent $\sum T_{\rm rise}$ (corresponding to the left y-axis), the triangles represent $\sum T_{\rm inter}$ (corresponding to the right y-axis), and the dash-dot lines show the linear fitting results.
To clearly present the data sets, we shifted the $\sum{\Delta\nu_{\rm max}}$ and $\sum T_{\rm inter}$ relationship to the right by 10 nHz. 
Additionally, the parameters for slow glitches 1–7 are derived from \cite{Shabanova2010}, but unfortunately, the $T_{\rm start}$ and $T_{\rm rise}$ data for slow glitch 3 were not provided.
\label{glitch:sdfo}} 
\end{figure*}  

To further validate the change in the modulation period of $\dot{\nu}$, we re-analyzed the oscillations before and after MJD 56716 using the autocorrelation function, with error bars calculated using the approach of \cite{PereraSWL2015}. The autocorrelation function results, shown in Fig.~\ref{0922:acf}, indicate that the modulation periods before and after MJD 56716 are approximately 560 days and 600 days, respectively. 
These results are generally consistent with the Lomb-Scargle analysis. 
Both methods provide period estimates with considerable uncertainty, but together offer evidence favoring a change in the modulation period.

\cite{PereraSWL2015} first noted that the modulation period of $\dot{\nu}$ in PSR J0922+0638 fluctuated after MJD 52000, decreasing from 630 days to 550 days, and speculated that this change occurred gradually. Subsequently, \cite{ShawSWB2022} used Gaussian Process Regression (GPR) to predict $\dot{\nu}$, and found that the modulation period decreased by 60 days after MJD 52000. This supports the idea of a gradual shortening of the modulation period.

Our Lomb-Scargle analysis indicated that before MJD 56716, the modulation period of $\dot{\nu}$ was approximately 537(24) days, consistent with previous reports. However, after MJD 56716, the modulation period increased to about 600(58) days.

Unfortunately, due to a lack of observational data during the transition from a 537(24) day to a 600(58) day modulation period, the specific cause of this change cannot be definitively determined. However, we note that \cite{ShawSWB2022} reported an abnormal three-peak fluctuation in $\dot{\nu}$ (with a period of about 800 days) after MJD 56500, which eventually returned to normal. We speculate that the change in the modulation period after MJD 56716 could be related to this abnormal fluctuation, supporting \cite{PereraSWL2015}'s view that the change was gradual.
Our detailed analysis of the long-term evolution of $\dot{\nu}$ in several other pulsars is currently in progress.

\section{Discussion} \label{sec:Discussion}

Slow glitches, a rare type of glitch, are characterized by a gradual increase in the pulsar’s spin frequency over hundreds of days, accompanied by a sudden jump in the first derivative of the spin frequency. The first slow glitch was reported in PSR J1825$-$0935 (B1822$-$09) by \cite{Shabanova1998} and \cite{ZouWWM2004}. To date, 31 slow glitch events have been identified across seven pulsars \citep{Shabanova1998, ZouWWM2004, ShabanovaT2007, ShabanovaT2009, Shabanova2010, YuMHJ2013, ZhouZZL2019}.

During 22 years of observing PSR J0922+0638 (2001--2023), we detected ten complete slow glitch events. In addition, using data from the Nanshan telescope, we identified more than 20 new slow glitch events in other pulsars, which will be discussed in a forthcoming paper.

The physical mechanisms behind slow glitches remain unclear. Observational data reveal variability in slow glitch behavior across pulsars. Some pulsars experience only a single slow glitch \citep{ZhouZZL2019}, while PSR J0922+0638 exhibits multiple slow glitches, accompanied by periodic oscillations in spin frequency over long timescales. The superfluid vortex model suggests that glitches occur when vortices in the neutron star’s superfluid unpin, transferring angular momentum to the solid crust, leading to a spin increase \citep{AlparCP1989}. Given the linear correlation between $\sum{\Delta \nu_{\rm max}}$, $\sum{T_{\rm inter}}$, and $\sum{T_{\rm rise}}$, the slow glitches in PSR J0922+0638 may originate from a similar superfluid process.
Another hypothesis, proposed by \cite{2007whsn.conf..153I}, suggests that slow glitches result from the solid crust’s delayed response to spin-down, which causes misalignment between the symmetry axis and the rotation axis. Once the accumulated stress exceeds a threshold, the crust realigns these axes, producing a slow glitch. Similarly, \cite{2008MNRAS.384.1034P} explored a mechanism where crustal fractures, under critical stress, temporarily transform surface material into a highly viscous fluid. This dissipates accumulated elastic and gravitational potential energy, driving slow glitches. However, further studies are needed to support this hypothesis.

For PSR J0922+0638, the ten slow glitches occurred quasi-periodically, with an average interval of approximately 553(21) days. As shown in Fig. \ref{glitch:sl}, these slow glitches significantly affect the evolution of $\dot{\nu}$, as detailed in Sec. \ref{Slow:glitches}. Lomb-Scargle analysis suggests that the modulation periods of $\dot{\nu}$ before and after MJD 56716 were approximately 537(24) days and 600(58) days, respectively, aligning closely with the average interval between slow glitches. Thus, we hypothesize that slow glitches may contribute to the periodic modulation of $\dot{\nu}$, consistent with \cite{Shabanova2010}.

In Fig. \ref{glitch:sdfo}, we explore the potential correlation between the jump amplitude $\Delta \nu_{\rm max}$ of the slow glitches and other parameters. Panels (a)--(c) show no significant correlation between $\Delta \nu_{\rm max}$ and the glitch start epoch $T_{\rm start}$, rise time $T_{\rm rise}$, or the interval time $T_{\rm inter}$. Notably, slow glitch 17 deviates from the typical pattern, with larger values of $\Delta \nu_{\rm max}$, $T_{\rm rise}$, and $T_{\rm inter}$ compared to the other glitches, suggesting a change in the characteristics of subsequent slow glitches. In panel (c), another outlier is seen for slow glitch 12, where glitch 2 interfered with the recovery process, resulting in the shortest glitch interval, $T_{\rm inter} \sim$ 343(7) days.

Interestingly, in panel (d) of Fig.~\ref{glitch:sdfo}, we observed clear linear correlations between $\sum{\Delta \nu_{\rm max}}$ and $\sum{T_{\rm rise}}$, as well as between $\sum{\Delta \nu_{\rm max}}$ and $\sum{T_{\rm inter}}$, confirmed through linear fitting (grey curve). The last data point of $\sum{T_{\rm inter}}$ deviates from the linear trend, which we attribute to the underestimated $T_{\rm inter}$ value of slow glitch 17, as the timing data ends at MJD $\sim$ 60033, missing the full event.

For pulsars with periodic oscillations in spin-down rate, some researchers attribute this to free precession \citep{2006MNRAS.365..653A}. However, this explanation struggles to account for the observed correlation between spin-down rate variations and pulse profile changes in certain pulsars \citep{stairs2019}. In fact, many pulsars show a connection between pulse profile evolution and spin-down rate fluctuations. \cite{LyneHSS2010} proposed that this behavior results from periodic switching between two stable magnetospheric states. Given that PSR J0922+0638 also shows spin-down rate variations accompanied by pulse profile changes, a similar mechanism may be at play, with the pulsar’s magnetosphere periodically switching between two states over hundreds of days. Abrupt transitions could lead to sudden spin-down rate changes (slow glitches), while long-term periodic switching could manifest as quasi-periodic spin-down rate oscillations. Nonetheless, further investigation is required to determine the dominant physical mechanism behind the spin-down behavior of PSR J0922+0638. Additionally, the long-term evolution of the pulsar may be influenced by a combination of superfluid dynamics, free precession, and magnetospheric state switching. 

\section{Conclusions} \label{sec:Conclusions}
In this work, we analyzed the rotational behavior of PSR J0922+0638, focusing on its spin-down behavior, timing residuals, and glitch events, using timing data from the Nanshan radio telescope collected over nearly 21 years (2001--2022). Additionally, over three years of data from the MeerKAT telescope (2020--2023) were incorporated to improve the accuracy of our analysis.

Over the 22-year observation period, we first examined normal glitch events and successfully detected two glitches, including the previously reported large glitch (glitch 2). We also identified a previously unreported small glitch (glitch 1) at MJD $\sim$ 53325(3), with a fractional frequency jump of $\Delta \nu/\nu \sim 0.79(6) \times 10^{-9}$.

We observed significant red noise in the timing residuals, and an analysis of the timing noise revealed a noticeable change in the spectral index before and after the large glitch (glitch 2), suggesting that the glitch events had a clear impact on the pulsar’s spin behavior.

Our observations also revealed notable modulation in the pulsar's spin, including multiple slow glitches and periodic changes in the spin-down rate, providing insights into the mechanisms driving these variations. Specifically, we identified ten slow glitch events, half of which were newly detected. The maximum amplitude of the spin frequency change for these events ranged from $\Delta{\nu_{\rm max}} \sim$ 2.61--9.47 nHz, with rise times between $T_{\rm rise} \sim$ 184--347 days. 
These glitches occurred quasi-periodically, possibly with an average interval of around 553(21) days.

Additionally, we found that $\dot{\nu}$ fluctuated periodically, with possible periods of about 537(24) days before the 700-day data gap at MJD 56716, and 600(58) days afterward, likely due to spin-down noise. These periodic oscillations in $\dot{\nu}$ align closely with the intervals between slow glitches, suggesting that slow glitches play an important role in the pulsar’s long-term spin behavior.

Based on the observed correlations between pulse profile changes and variations in the spin-down rate, we propose that the pulsar’s magnetosphere may periodically switch between two stable states, which could contribute to both slow glitches and the periodic oscillations in $\dot{\nu}$.

Our study provides new insights into the nature of slow glitches and their impact on the spin behavior of PSR J0922+0638, suggesting complex interactions between the pulsar’s internal dynamics, magnetosphere, and long-term evolution. Continued high-cadence, high-precision observations will be essential for deepening our understanding of pulsar magnetospheres and their internal physics.

\section*{Acknowledgments}
We appreciate discussions with Z. Yan, S. J. Dang, S. Q. Zhou, E. G{\"u}gercino{\u{g}}lu and the support of TianMa Telescope Team from the Shanghai Astronomical Observatory.
The work is supported by the National SKA Program of China (No.~2020SKA0120300), the Strategic Priority Research Program of the Chinese Academy of Sciences (No. XDB0550300), the National Natural Science Foundation of China (Nos. 12273028, 12494572 and~12041304), the Major Science and Technology Program of Xinjiang Uygur Autonomous Region (No. 2022A03013, 2022A03013-4), the Natural Science Foundation of Xinjiang Uygur Autonomous Region (No. 2023D01E20) and XMU Training Program of Innovation and Enterpreneurship for Undergraduates.
The Nanshan 26 m Radio Telescope is partly supported by the Operation, Maintenance and Upgrading Fund for Astronomical Telescopes and Facility Instruments, budgeted from the Ministry of Finance of China (MOF) and administrated by the Chinese Academy of Sciences (CAS).
The MeerKAT telescope is operated by the South African Radio Astronomy Observatory (SARAO), which is a facility of the National Research Foundation, an agency of the Department of Science and Innovation. 
SARAO acknowledges the ongoing advice and calibration of GPS systems by the National Metrology Institute of South Africa (NMISA) and the time space reference systems department department of the Paris Observatory. 

\vspace{5mm}
\facilities{Nanshan 26-m radio telescope, MeerKAT radio telescope}

\software{\texttt{TEMPO2} \citep{EdwardsHM2006,HobbsEM2006},  
          \texttt{PSRCHIVE} \citep{HotanSM2004,StratenDO2012}, 
          }

\bibliography{J0922}{}
\bibliographystyle{aasjournal}

\end{document}